\documentclass[10pt,conference]{IEEEtran}

\usepackage{cite}
\usepackage{amsmath,amssymb,amsfonts}
\usepackage{algorithmic}
\usepackage{graphicx}
\usepackage{textcomp}
\usepackage{times}
\usepackage{epsfig}
\usepackage{placeins}
\usepackage{amsmath}
\usepackage{amssymb}
\usepackage{listings}
\usepackage{xcolor}
\usepackage{tabularx}
\usepackage{adjustbox}
\usepackage{spverbatim}
\usepackage{xcolor,colortbl}
\usepackage{booktabs}

\usepackage{caption} 
\usepackage[bookmarks=false,hidelinks]{hyperref}

\begin{document}
\begin{sloppy}

\author{\IEEEauthorblockN{Matthew Kelly}
\IEEEauthorblockA{Cohda Wireless\\
matt.kelly@cohdawireless.com} \and
\IEEEauthorblockN{Christoph Treude}
\IEEEauthorblockA{University of Adelaide\\
christoph.treude@adelaide.edu.au} \and
\IEEEauthorblockN{Alex Murray}
\IEEEauthorblockA{Canonical\\
alex.murray@canonical.com}
}

\title{A Case Study on Automated Fuzz Target Generation for Large Codebases}

\IEEEoverridecommandlockouts
\IEEEpubid{\makebox[\columnwidth]{978-1-7281-2968-6/19/\$31.00~\copyright2019 IEEE \hfill} \hspace{\columnsep}\makebox[\columnwidth]{ }}
\maketitle
\IEEEpubidadjcol

\maketitle

\begin{abstract}
Fuzz Testing is a largely automated testing technique that provides random and unexpected input to a program in attempt to trigger failure conditions. Much of the research conducted thus far into Fuzz Testing has focused on developing improvements to available Fuzz Testing tools and frameworks in order to improve efficiency. In this paper however, we instead look at a way in which we can reduce the amount of developer time required to integrate Fuzz Testing to help maintain an existing codebase. We accomplish this with a new technique for automatically generating Fuzz Targets, the modified versions of programs on which Fuzz Testing tools operate. We evaluated three different Fuzz Testing solutions on the codebase of our industry partner and found a fully automated solution to result in significantly more bugs found with respect to the developer time required to implement said solution. Our research is an important step towards increasing the prevalence of Fuzz Testing by making it simpler to integrate a Fuzz Testing solution for maintaining an existing codebase.
\end{abstract}

\section{Introduction}

It is nearly impossible to write flawless code for anything but the simplest of problems. During any development process, mistakes are made, and bugs get introduced into the code. These bugs can be anything from simple, unintended but benign behaviour, to serious security vulnerabilities that an attacker could use to gain control of the entire system upon which the software is being run. It is because of this that testing is one of the most vital components of software development, and on large scale codebases, finding ways of automating the testing process is invaluable~\cite{Liang}.

Fuzz Testing or Fuzzing is a specific type of automated testing that provides invalid or unexpected data as an input to a target program and then monitors for any exceptions or crashes which that input produces. One often used method of automated Fuzz Testing is through the use of a coverage guided tool such as libFuzzer or American Fuzzy Lop (AFL). These two tools take an input corpus and, using genetic algorithms~\cite{Mitchell}, apply random mutations to that corpus selecting for inputs that find new paths through the code. Between them, these tools have found thousands of bugs and vulnerabilities in many commonly used applications and libraries~\cite{oss-trophies, libFuzzer, afl-trophies}.

These tools are a useful addition to any test suite, but most programs cannot be immediately fuzzed, `as is'. In most cases it is necessary to build a Fuzz Target, a modified version of the program designed to work with the selected fuzzing tool~\cite{libFuzzer}. In the case of AFL, such a Fuzz Target is simply a program that reads an input from stdin, performs some operations on that input, and then exits. For libFuzzer, a specific function needs to be implemented that takes an input, performs some operations on it, and then cleans up after itself. Using either of these two tools, or anything similar, will require a developer to spend time creating these Fuzz Targets. Conventional wisdom is to keep Fuzz Targets small and fuzz single pieces of functionality at a time. For instance, to Fuzz Test a tool that accepts input in several different formats, it would likely be most efficient to build an individual Fuzz Target for each format~\cite{libFuzzer}. This is because this method reduces the input space for each target, causing each random mutation of the fuzzing tool to have a better chance of finding a new code path. For small scale projects, creating these Fuzz Targets may not involve much work, as there is potentially as few as one Fuzz Target that would need to be made. However, for large codebases there could potentially be hundreds of different locations in the code that could benefit from being Fuzz Tested. Also, in order to create a Fuzz Target, the developer needs to have a firm understanding of both the code for which they are building the target and the Fuzz Testing framework they are using. Realistically, an organisation with an existing codebase, that wishes to add Fuzz Testing to its existing test suite, will need to spend extra time teaching developers how to build these Fuzz Targets. 

In this paper we explore automated generation of Fuzz Targets as a way to reduce the barrier to entry for setting up Fuzz Testing to maintain an existing codebase, in an attempt to incentivise companies to implement additional testing practices. We developed a tool to generate a libFuzzer compatible Fuzz Target based on a set of given function signatures in a piece of C code, and compare the performance of those Fuzz Targets to that of more conventional fuzzing methodologies as described below. The tool developed can be used on any codebase written in C, but will be most effective in comparison to traditional Fuzz Testing methods on larger codebases for which a comprehensive set of Fuzz Targets could not quickly be created manually.

The codebase used in this work was the protocol stack of our industry partner Cohda Wireless. Cohda Wireless is a global leader in the development of Connected Autonomous Vehicle software with proven applications for Smart City, Mining and other environments. Cohda is headquartered in Australia and has offices in Europe, China, and the USA. \mbox{Cohda} Wireless's innovative software solutions enable connected vehicles and connected autonomous vehicles to communicate with other vehicles and with Smart City infrastructure. These connections span Vehicle-to-Vehicle, Vehicle-to-Infrastructure, and Vehicle-to-Pedestrian (collectively called V2X), allowing CAVs to `talk' to each other, Smart Cities, and vulnerable road users in order to avoid accidents, reduce congestion, and be more efficient. Because of this, much of the code in this stack is safety critical, increasing the necessity to ensure that it is adequately tested. At the time of writing, the stack contains 300,000 lines of code and 17,000 functions. It utilises a continuous integration pipeline, regularly running a comprehensive suite of unit and application tests, to ensure the continued reliability of the code, and to prevent regression.

Because of the size of this codebase, retroactively adding comprehensive Fuzz Testing would require a considerable amount of effort and resources. This makes the codebase an ideal candidate for this work, as it would be of considerable benefit, both in terms of developer time and code quality, if the addition of Fuzz Testing could be largely automated.

We tested two different approaches utilising Fuzz Targets generated in a semi-automated and fully automated manner against the standard approach of building Fuzz Targets manually. We found both of these approaches to result in significantly more bugs found for the same amount of developer time required to implement them.

\section{Literature Review}

Fuzz Testing can trace its origins to as early as the 1950s when it was common practice to test programs using decks of punch cards chosen at random or recovered from the trash in order to find undesired behaviours~\cite{fuzz-history}. For some time this type of testing was thought to be an ineffective means of testing programs until the practice of testing programs with random inputs was formally investigated by Duran and Ntafos~\cite{Duran}, who found random testing to be a viable and cost effective testing practice.

There has been a lot of research into fuzzing from a theoretical standpoint. This research has been primarily in the development of new and effective tools, or optimisation of the performance of currently existing tools. Such developments include AFL by Zalewski~\cite{afl}, a fuzzer which employs genetic algorithms to efficiently increase code coverage of test cases. Research into optimisation includes work by Rebert et al.~\cite{Rebert}, who investigate optimising the seed corpus for the fuzzer in order to maximise the total number of bugs found over a fixed timeframe. Another work is from Xu et al.~\cite{Xu} that focuses on developing improved operating primitives that decrease the execution time of each iteration of the fuzzer.  

While they are valuable contributions, a perspective on Fuzz Testing research that is not usually taken is how best to apply Fuzz Testing techniques to existing codebases, particularly when these codebases are large, as would be found in an industry context. There is some work being done with respect to Continuous Fuzzing, the process of running Fuzz Tests either periodically, when new revisions are committed to version control, or some combination of both. An example of this is Google's OSS-Fuzz~\cite{oss-fuzz}, a continuous fuzzing framework for open-source software. OSS-Fuzz takes any open source software that is widely used, or part of core infrastructure, and, using Fuzz Targets created by each open source project's maintainers, runs Fuzz Tests on ClusterFuzz, Google's distributed fuzzing infrastructure~\cite{cluster-fuzz}. 

\section{Target Generator}

To enable our industry partner to integrate a Fuzz Testing solution into their existing testing framework, we developed Fuzz Target Generator (FTG), a tool for automatically generating Fuzz Targets to be used with libFuzzer. Unlike existing tools, Fuzz Target Generator is capable of producing Fuzz Test ready code directly from an existing codebase without the need for additional input from a developer. We describe the design of Fuzz Target Generator in the following paragraphs.

To build a Fuzz Target for libFuzzer in C, a developer needs to implement the function in Listing \ref{testOne-C} that performs some operation on an input array and link the libFuzzer library which contains a main function that will call the target repeatedly with each new input generated from the mutated seed corpus.

\begin{lstlisting}[caption={libFuzzer input function for C}, captionpos=b, label=testOne-C]
int LLVMFuzzerTestOneInput(uint8_t *data,  
                           size_t size)
\end{lstlisting}

The primary challenge is serialising the parameter list of the function so that all the necessary data to call the function can be mapped to a single contiguous input buffer. This allows the array of \verb’uint8_t’ passed to the Fuzz Target by libFuzzer to be split up into all the necessary variables needed to call the target function. This should be a reasonably simple task for any function signature containing only basic types or structures containing only basic types. Function signatures containing arrays should also be possible if it can be indicated to the tool which value is the array and which is its corresponding length. More complex structures however, such as structures containing pointers to other structures, are more difficult to serialise. While possible in most cases, such structures are not considered in this work yet.

In the following, we describe the cases which FTG can handle in more detail. First, consider the function signature containing a structure in Listing \ref{ex1}.

\begin{lstlisting}[caption={Function signature containing structure}, captionpos=b, label=ex1]
struct Foo {
	int x;
	char y;
	float z;
}
//@fuzztest
void func1(int a, int b, struct Foo f) { 
	...
}
\end{lstlisting}

\begin{figure}[h]
\begin{center}
\includegraphics[width=.5\linewidth]{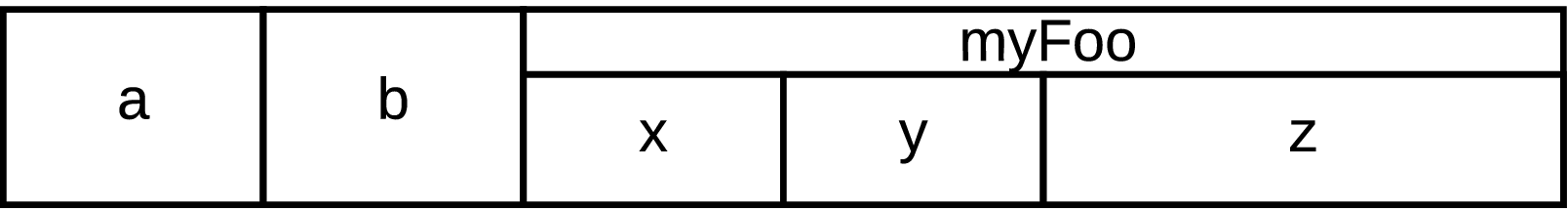}
\end{center}
   \caption{Serialised input for func1}
\label{fig:func1}
\end{figure}

The parameter list can be serialised with relative ease as is shown in Figure \ref{fig:func1}, as all of the parameters are simple types or structures containing only simple types, and thus, calling \verb'sizeof' will return the exact number of bytes that need to be allocated for that type in the input format. The commented line above the function declaration contains an \verb’@fuzztest’ directive, which is used to indicate to the target generator that a Fuzz Target should be generated for the following function. This directive is part of FTG and will be described in detail in the next section. Next consider the function in Listing \ref{func2} containing an array and its length.

\begin{lstlisting}[caption={Function call with array}, captionpos=b, label=func2]
//@fuzztest Array(arr, len)
void func2(int *arr, int len) {
	...
}
\end{lstlisting}

Using the same approach as above, this signature would be serialised as shown in Figure \ref{fig:naive-func2}.

\begin{figure}[h]
\begin{center}
\includegraphics[width=0.3\linewidth]{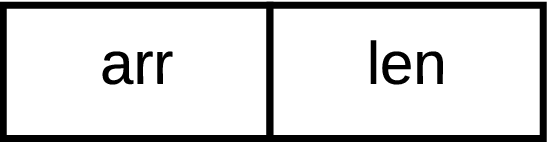}
\end{center}
   \caption{Naive serialised input for func2}
   \label{fig:naive-func2}
\end{figure}

We would be naive to serialise func2 in this way as it would result in an array of size one and a length determined by the input data from libFuzzer being passed to the target function. A Fuzz Target reading inputs in this format will not be very effective as it is unlikely that a function that takes an array will have much of its functionality tested when only arrays of size one are passed. It is also possible that false positive bugs could be found due to the array size not matching the actual length of the array passed. However, if we were able to indicate to the tool that these two values should instead be considered an array and its length, such as by using an additional Array parameter in the \verb'@fuzztest' directive in the snippet above, we could instead serialise the signature as shown in Figure \ref{fig:accurate-func2}.

\begin{figure}[h]
\begin{center}
\includegraphics[width=.5\linewidth]{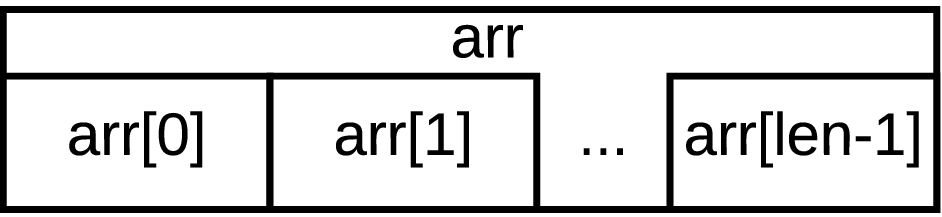}
\end{center}
   \caption{Accurate serialised input for func2}
   \label{fig:accurate-func2}
\end{figure}

A Fuzz Target reading inputs in this format is going to be able to consider a much wider range of inputs and as such, is likely to be much more effective. This is because arrays of lengths other than one can be passed to the target function and the array length passed will be accurate.

Once we have created an input format that can be mapped to the function signature, it is simply a matter of generating a Fuzz Target that can separate out each part of the input into a correctly typed variable, and then pass all those variables to the function being tested. However, this is more complicated than it may seem. With libFuzzer the input is treated as an array of \verb'uint8_t', a type with a size of one byte, thus for any type of a larger size, multiple values from the input array need to be treated as the same value when being passed to the target function. libFuzzer Fuzz Targets for the first function described above could look as shown in Listing \ref{func1-target}.

\begin{lstlisting} [caption={Generated Fuzz Target for func1}, captionpos=b, label=func1-target]
int LLVMFuzzerTestOneInput(uint8_t *data,  
                            size_t size)
	int a = 0;
	char b = 0;
	struct Foo myFoo;
	uint8_t *pos = data;
	memcpy(&a, pos, sizeof(int));
	pos += sizeof(int);
	memcpy(&b, pos, sizeof(char));
	pos += sizeof(char);
	memcpy(&myFoo, pos, sizeof(Foo));
	func1(a, b, myFoo);
	return 0;
}
\end{lstlisting}

This code utilises \texttt{memcpy} in order to map the smaller values of the \verb'uint8_t' array onto the larger values of \verb'int' and \verb'Foo'.

\section{Tool Usage}

To mark a selected function to have a Fuzz Target generated a user needs to add the \verb'@fuzztest' directive in a commented line directly above the function. By default this will generate a Fuzz Target that allocates an amount of data from the libFuzzer input buffer to each parameter equal to the size of that parameter's type. This behaviour can be modified for specific parameters using the additional directives listed below.

\subsection{Array(array\_ptr, array\_len)}

This directive indicates two values that should be treated as an array, rather than individual parameters as they would by default. The array is placed at the end of the fuzzer input format, and the Fuzz Target will pass any data remaining in the input buffer after filling all the other parameters as the array. Because of this, only one Array can currently be supported in a single Fuzz Target.

\begin{itemize}
	\item \verb'array_ptr': pointer to the start of the array
	\item \verb'array_len': parameter indicating the length of the array
\end{itemize}

\textbf{Example}

\begin{lstlisting}
//@fuzztest Array(pBuf, Len)
void foo(char * pBuf, int Len) {
	...
}
\end{lstlisting}

\subsection{Value(parameter, value)}

This directive indicates to use the same value for a given parameter every time, instead of allowing it to be chosen by the fuzzer. This is useful for disabling optional features such as error pointers or to fix the value of flags or enums passed to the target function. Fixing values allows more targeted control over which code paths are explored, and reduces the required input size, both of which increase the efficiency of Fuzz Testing.

\begin{itemize}
	\item \verb'parameter': parameter to be set
	\item \verb'value': value for the parameter to be set to
\end{itemize}

\textbf{Example}

\begin{lstlisting}
//@fuzztest Value(e, ENUM_VALUE)
//Value(pErr, NULL)
void foo(int a, myEnum e, int * pErr) {
	...
}
\end{lstlisting}

\subsection{Output(parameter)}

This directive marks a parameter as being used for output. This will cause a variable to be initialised and passed to the function being called, but will not initialise it using data provided by the fuzzer. This reduces the required size of the input, as if the parameter is never going to be read by the target function, any data used to initialise it would be wasted.
\begin{itemize}
	\item \verb'parameter': parameter to be marked as output
\end{itemize}

\subsection{Cleanup(condition, function [, params])}

This directive indicates to call a function to help clean up the fuzzing run after the target function has been called. This is most useful if the target function allocates memory that needs to be freed by a separate function. An optional condition is also available that will be evaluated to determine if the clean-up function should be called. The special name \verb'fuzzer_return_value' can be used as part of the clean-up to get the value returned by the target function. This is useful as it prevents the Fuzz Target from creating false positive memory leaks, as well as potentially obscuring real bugs in the target function.

\begin{itemize}
	\item \verb'condition': the condition to evaluate before calling the clean-up function, can be empty
	\item \verb'function': the clean-up function to be called
	\item \verb'params': a list of parameters to be passed to the clean-up function
\end{itemize}

\textbf{Examples}

\begin{lstlisting}
//@fuzztest 
//Cleanup(, free, fuzzer_return_value)
char * allocate_some_memory(
	int this_much) {
	...
}
//@fuzztest Output(out) 
//Cleanup(out != NULL, myTypeFree, out)
void maybe_allocate_memory_to_output(
	int this_much, myType * out) {
	...
 }
\end{lstlisting}


\section{Case Study}

Once we were able to generate these Fuzz Targets, we evaluated their performance and compared them against the performance of a Fuzz Target that was hand-made by a developer and utilises a quality seed corpus. This was done as part of the case study with our industry partner. 

For this evaluation, the performance metric we used was the number of bugs found over a fixed period of time, namely ten hours. We ran three different tests on the J2735 Decoding module of our industry partner's stack, on the same revision within the version control system.

The three tested approaches are as follows, including the number of Fuzz Targets created by that method, and the amount of developer time used.

\begin{itemize}
\item \textbf{Automated}: 66 Fuzz Targets generated from every function for which the tool was able, without any additional input from a developer. 

\textit{Effort}: This method took no additional time to set up, outside of running the target generator and the ten hour libFuzzer run time.
\item \textbf{Annotated}: Six Fuzz Targets generated from functions specifically marked with the \verb'@fuzztest' directive and any additional information, such as the Array parameter described above, by a developer familiar with the codebase, who was provided with the documentation listed under the Tool Usage section above, on their discretion of which functionality would benefit most from Fuzz Testing. 

\textit{Effort}: This method took nominally five minutes to select and annotate the chosen functions.
\item \textbf{Handmade}: A single handmade Fuzz Target with included seed corpus was run for the full ten hours. 

\textit{Effort}: This target took approximately one and a half hours to create and gather the corpus for.

\end{itemize}

For the first and second test each Fuzz Target was run using round-robin scheduling, which splits up the allotted time equally between each target. Given a total number of Fuzz Targets \(n\) and an allotted time period of \(T\) each target is run for a total time of \(T/n\). All experiments were run on an Ubuntu 16.04 VM with a 2.6GHz Quad Core Intel i7-6600U and six GB of RAM. All code was built using Clang 6.0 with AddressSanatizer. This allows the Fuzz Targets to detect memory issues as well as crashes. AddressSanatizer also has a feature that produces deduplication tokens in its output that can be activated by setting the environment variable \verb'ASAN_OPTIONS=dedup_token_length=3'. In some cases there could be a Fuzz Target that is capable of taking a codepath into another function that has its own Fuzz Target. This could result in two different Fuzz Targets finding the exact same bug, but because AddressSanatizer's deduplication tokens are based on the stack trace, this will allow it to handle the case in which two Fuzz Targets find the same bug. It is important to note that in order to produce deduplication tokens, AddressSanatizer needs to be able to symbolise its stack traces. When building with Clang, this means that AddressSanatizer needs to know where to find the llvm-symbolizer binary corresponding to the version of Clang used. This can either be done by adding it to \verb'$PATH' or by setting \verb'ASAN_OPTIONS=external_symbolizer_path=' \verb'/path/to/llvm-symbolizer'. If this is not done, it will be much more difficult to deduplicate any bugs found by libFuzzer.

Once each test was run, all of the found bugs were verified by re-running them with the appropriate Fuzz Target to ensure that a crash or leak was replicable. The AddressSanitizer deduplication tokens described above for each bug found were inspected to ensure that there were no duplicates. As libFuzzer is non-deterministic, each test was then repeated five times and the total number of bugs found in each iteration were averaged.

We ran our experiments on a single module in the stack of our industry partner, the J2735 Decoding module. J2735 is a SAE (Society of Automotive Engineers) standard message set for use primarily by applications utilising the 5.9GHz Dedicated Short Range Communications (DSRC) for Wireless Access in Vehicular Environments (WAVE) communications systems~\cite{j2735}. It is the wireless message set used by all of our industry partner's applications built for North American end users, thus making the J2735 decoding module core infrastructure within the stack.

To illustrate the methodology of our case study, we present an example of a Fuzz Target created with each of the three methods for the same target function in our online appendix at \url{http://doi.org/10.5281/zenodo.3345339}.

\begin{figure}[b]
\vspace{-.5cm}
\begin{center}
\includegraphics[width=0.8\linewidth]{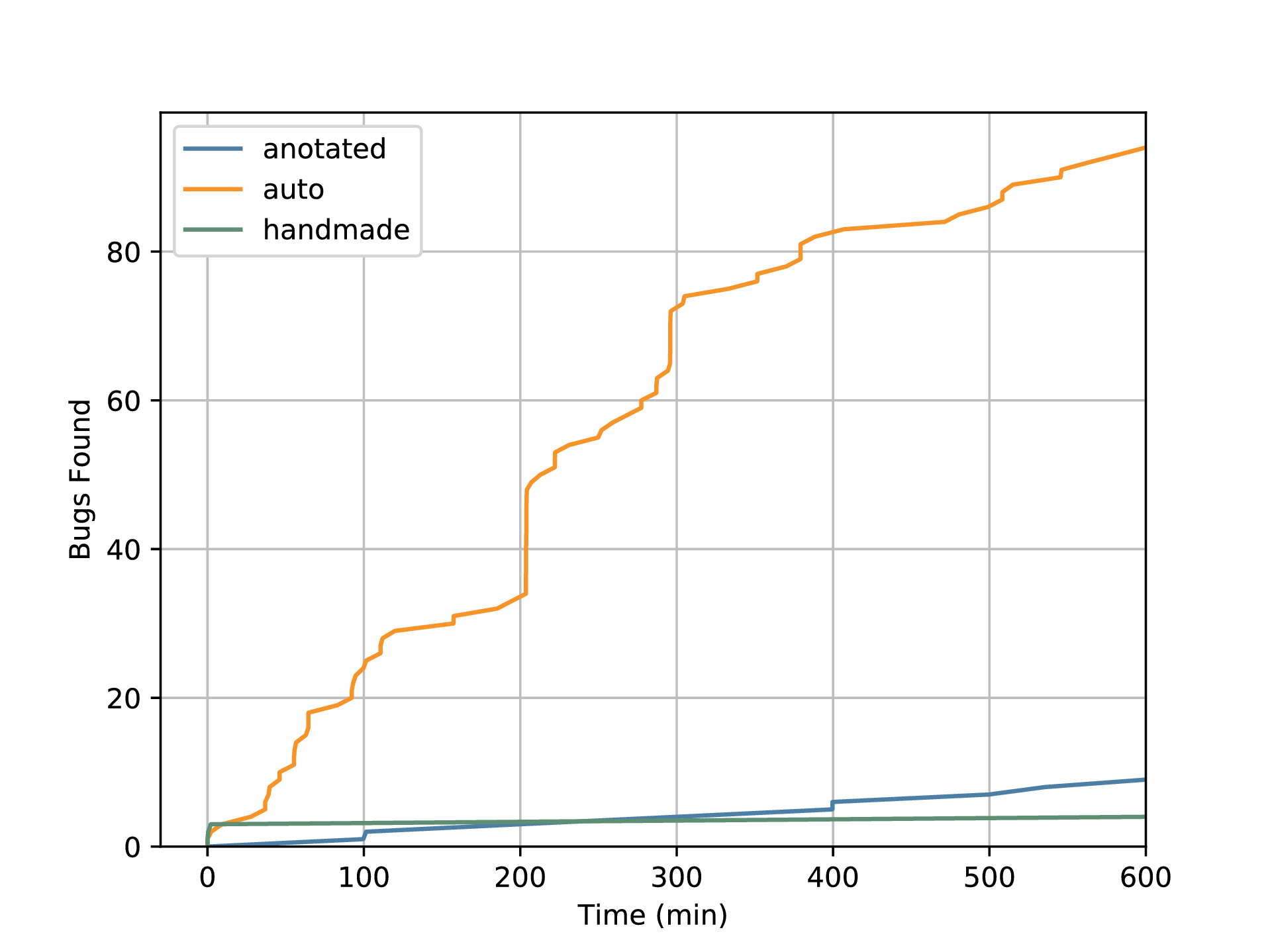}
\end{center}
\vspace{-.5cm}
   \caption{Bugs over time}
   \label{resultsGraph}
\end{figure}

\begin{table}[t]
\centering
\caption{Results (averaged across five runs)}
\begin{tabular}{l@{}rrrrrr}
\toprule
Run & Targets & Distinct & Unique & Unique & Total & Bugs/ \\
& & Bug Inputs & Crashes & Leaks & Bugs & Target \\
\midrule
Automated & 66 & 209,797.6 & 82.6 & 11 & 93.6 & 1.42 \\
Annotated & 6 & 82,258.8 & 6.8 & 3 & 9.8 & 1.66 \\
Manual & 1 &  73,465.0 &  1.8 &  3 &  4.8 &  4.8 \\
\bottomrule
\end{tabular}
\label{resultsTable}
\end{table}

\section{Results}

In this section, we describe the results of our case study, separately for the total number of bugs found by each approach and the bugs found per target.

\paragraph{Total Bugs}
The results presented in Table~\ref{resultsTable} show that the fully automated method of generating Fuzz Targets found the most bugs by a large margin, finding nearly ten times more bugs than the annotated method and twenty times more bugs than the manually created Fuzz Target. Not only does this method result in the most bugs found, it also takes the least developer time to implement. 

Consider also Figure~\ref{resultsGraph}, showing the number of bugs found over time for each of the three methods. Most bugs tend to be found shortly after each Fuzz Target begins running. This is indicative of `shallow bugs', i.e., bugs that can be found almost right away by Fuzz Testing as they occur on a large range of inputs. `Shallow bugs' tend to obscure other bugs because when a bug is found by libFuzzer, the process is stopped~\cite{libFuzzer-tute}. It is possible to mitigate this by restarting the Fuzz Target when a bug is found, but this requires initialisation to be re-performed. Thus, if a large number of the inputs tried by libFuzzer cause bugs, the efficiency of the tool will be decreased. However, this is not a limitation of the generated Fuzz Targets, but rather of libFuzzer, the current primary mitigation for which is fixing the offending bug to allow Fuzz Testing to progress. 

\paragraph{Bugs per Target}

As most bugs were found shortly after a Fuzz Target began running, the amount of time each target ran for is less of a factor in how many bugs were found. It is because of this that the relative quality of Fuzz Targets created by each of the three methods can be estimated by the number of bugs found per target. As Table \ref{resultsTable} shows, the handmade Fuzz Target with corpus produces the most bugs with an average of 4.8 bugs found per run and the fully automated method finds the least bugs per target at an average of just 1.42 per run, which is to be expected as the annotated Fuzz Targets are typically of a higher quality as shown in the Fuzz Target example in the appendix. The annotated method produces slightly more bugs per target than the fully automated method, which would suggest that we would likely see an even greater number of bugs found if we would annotate all functions for which targets were generated in the fully automated method. Although annotating the functions appears to result only in a slight improvement in bugs found per target, the time required to perform the annotation is so minimal it is still worthwhile. 

\paragraph{False Positives}
To determine what proportion of bugs found were false positives, it was necessary to manually inspect a sample of the crashing inputs found. For the fully automated method, we inspected a sample of 15 of the 102 unique bugs found across all runs, as well as all the bugs found by both the annotated and handmade methods. 

In the fully automated method eleven out of the 15 unique bugs inspected were false positives due to flaws in the generated Fuzz Target. These flaws were primarily due to the target generator's inability to determine when two function parameters are an array and the corresponding length, i.e., the length parameter gets set to be longer than the data in the array, and out of bounds read or write can occur.

In the annotated method, only four of the 13 unique bugs found were false positives. However, all but one of those false positives could have been avoided with additional parameters in the \verb'@fuzztest' directive to better inform the target generator about how the target function operates.

For the handmade Fuzz Target, five unique bugs across all runs were reported, none of which were false positives. However, three of the reported bugs had very similar deduplication tokens and were caused by the same underlying bug.

\section{Discussion}

The fact that we have been able to generate a Fuzz Target for a large number of lower level functions, rather than creating fewer targets for higher level functions, means that we have reduced the complexity of the code paths that need to be taken to reach the same underlying bug. This likely means that nuanced bugs found by a high level Fuzz Target can be made shallow, and thus easier to find in our automated method as opposed to standard methods of creating Fuzz Targets.

The primary reason that false positive bugs could be introduced into the code by the generated Fuzz Targets is their ability to call a target function in a way that is not intended. This was clearly most prevalent in the fully automated target generation. False positive leaks also occurred due to the target function allocating memory without an appropriate Cleanup parameter being passed to the \verb’@fuzztest’ directive. It is also possible that the generated Fuzz Targets could find bugs that, although not false positives as described above, still could not occur during the normal operation of software. These `soft positives' could include functions that do not sanity check their own input, such as ensuring pointers are not null, values are within acceptable ranges, etc., instead relying on higher level functions to not pass them invalid data. In some cases, this could result in crashes or leaks being found that could not occur normally. However, `soft positives' are less of an issue than false positives, as they are still indicative of potential flaws in the code and have the possibility of causing errors in typical usage in the future as the codebase undergoes additional development.

\section{Limitations}

This case study considered only a single module within a single codebase and as such the results found may not generalise. The module used contained many functions that the tool was able to generate functional Fuzz Targets for using either the automated or annotated method, which may not be the case for all codebases or even for other modules within the codebase. Currently our tool can only handle function signatures with only basic types, or structures containing only basic types. 
Most of the bugs we found were `shallow bugs' found quickly after a target began running. 
Because of this, conducting the same experiments over a longer period of time could have impacted the results. We evaluate our experiments primarily using the number of unique bugs found. However, there is the possibility of the same bug being reported as multiple, or different bugs being reported as the same. We have taken care to deduplicate found bugs thoroughly, but we cannot say with certainty that all duplicates were detected without manual verification of all found bugs.

\section{Conclusion and Future Work}

In this work we evaluated a new method of automatically generating Fuzz Targets against the conventional method of creating them by hand. We did this in an attempt to reduce the amount of developer time required to integrate a Fuzz Testing solution to help maintain an existing codebase.

For meeting the needs of our industry partner, we consider the annotated method to be the most effective in terms of developer time to bugs found, when factoring in the time taken both to produce annotations and to separate false positives from true positives. Despite the potential for the fully automated method to find many more bugs, given the high false positive rate, it would take much less time to produce the annotations then it would to verify that the bugs found were valid and not due to flaws in the generated Fuzz Target. As long as sufficient time and care is taken in producing the annotations and ensuring that they are accurate, the false positive rate of the annotated method can be very low making it the most effective solution.

In future work, there are several potential improvements that could be made to the target generation tool in improving its efficiency, improving the function signature serialisation to handle more complicated structures, and adding to its ability to automatically determine how function parameters are used without the need for a developer to annotate them.

\end{sloppy}
\end{document}